\documentclass[epjCONF,columns]{svjour} 
\usepackage{graphics}
\usepackage[varg]{txfonts} 
\usepackage[latin1]{inputenc}

\usepackage{upgreek}
\usepackage{atlasphysics}

\session-title{2011 Hadron Collider Physics symposium (HCP-2011)}
\begin{document}
\title{Tau Lepton Reconstruction and Identification at ATLAS}
\author{Felix Friedrich\thanks{\email{felix.friedrich@tu-dresden.de}} \\ on behalf of the ATLAS Collaboration}
\institute{Institut für Kern- und Teilchenphysik, Technical University of Dresden, Dresden, Germany}
\abstract{
Tau leptons play an important role in the physics program at the LHC.
They are used in searches for new phenomena like the Higgs boson or
Supersymmetry and in electroweak measurements. Identifying hadronically 
decaying tau leptons with good performance is an essential part of these analyses.
We present the current status of the tau reconstruction and identification 
at the LHC with the ATLAS detector. The tau identification efficiencies and their 
systematic uncertainties are measured using $\Wtau$ and $\Ztau$ events, 
and compared with the predictions from Monte Carlo simulations. 
} 
\maketitle
\section{Introduction}
\label{intro}
Tau leptons are important signatures for Standard Model processes and searches for new physics. 
With a mass of $1.777 \GeV$, the tau is the heaviest lepton and due to its short lifetime of $2.9\times 10^{-13}\mathrm{s}$ $(c\tau=87\mathrm{\upmu m})$, the tau lepton decays inside the beam pipe of the LHC~\cite{LHCEvans:2008}. 
The tau lepton is the only lepton that has a hadronic decay mode. While it decays in 35\% of the time leptonically, the hadronic decay mode occurs 65\% of the time. The majority of hadronic tau decays are characterized by one or three charged pions usually accompanied by neutral pions.
The kinematics of QCD jets are similar to that of hadronically decaying $\uptau$ leptons, leading to a high potential probability for misidentifying them as tau leptons. In addition, the cross-section of most of the Standard Model and new physics processes with tau leptons in the final state are small compared to the overwhelming background from QCD processes at LHC. Therefore well performing tau identification is crucial.
In ATLAS~\cite{AtlasJinst:2008}, tau reconstruction and identification~\cite{ATLAS-CONF-2011-152} concentrates on the hadronic decay modes of a tau lepton. They are classified according to the number of reconstructed charged decay particles (prongs). These decays can be differentiated from QCD jets by their characteristics, such as low track multiplicity, collimated energy deposits, and in case of 3-prong tau leptons the displacement of the secondary vertex.
\section{Reconstruction}
\label{sec:1}
Calorimeter jets with a transverse energy larger than 10~\GeV\ and within the detector acceptance are used as a seed for the reconstruction of tau candidates. Tracks within a cone of $\mathrm{\Delta R}=\sqrt{(\mathrm{\Delta\phi})^2+(\mathrm{\Delta\eta})^2}<0.4$ around the tau axis passing certain quality criteria are associated to the tau candidate and used to calculate the discriminating variables. The number of tracks within $\mathrm{\Delta R}<0.2$ is used to classify the tau candidate into single- or multi-prong categories. Variables based on calorimeter information are calculated from calorimeter cells in $\mathrm{\Delta R}<0.4$ around the tau axis. The tau energy is calculated using all calorimeter clusters within a core of $\mathrm{\Delta R}<0.2$ around the 4-vector sum of clusters associated with the jet seed. Calibration factors are derived from response functions using Monte Carlo simulations, which come from the ratio of reconstructed tau energy to true visible tau energy. Response functions are functions dependent on the tau transverse momentum \pT, and calculated separately for single- and multi-prong tau leptons, as well as for different detector regions. The systematic uncertainties on the tau energy scale are fully derived from Monte Carlo and were found to be 4\%--7\%~\cite{ATLAS-CONF-2011-152}.
\section{Identification}
\label{sec:2}
\begin{figure}
\centering
\resizebox{0.75\columnwidth}{!}{ %
  \includegraphics{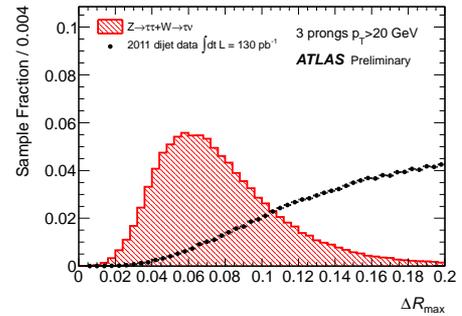} }
\caption{Maximal distance between a track and the tau axis, $\mathrm{\Delta R}_{\mathrm{max}}$. Only tracks inside a cone of $\mathrm{\Delta R}<0.2$ around the tau axis are considered~\cite{ATLAS-CONF-2011-152}.}
\label{fig:dRmax3p}       
\end{figure}

\begin{figure}
\centering
\resizebox{0.75\columnwidth}{!}{ %
  \includegraphics{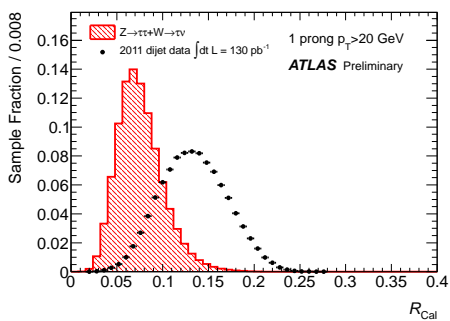} }
\caption{Energy weighted shower width in the calorimeter, $\mathrm{R}_{\mathrm{Cal}}$, for tau signal Monte Carlo (red) and compared to QCD di-jet data (black)~\cite{ATLAS-CONF-2011-152}.}
\label{fig:calRadius}       
\end{figure}

\begin{figure}
\centering
\resizebox{0.75\columnwidth}{!}{ %
  \includegraphics{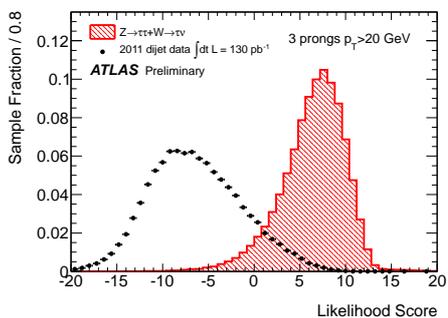} }
\caption{Output score of the projective likelihood tau identification method~\cite{ATLAS-CONF-2011-152}.}
\label{fig:Perf3pLLHscore}       
\end{figure}

\begin{figure}
\centering
\resizebox{0.75\columnwidth}{!}{ %
  \includegraphics{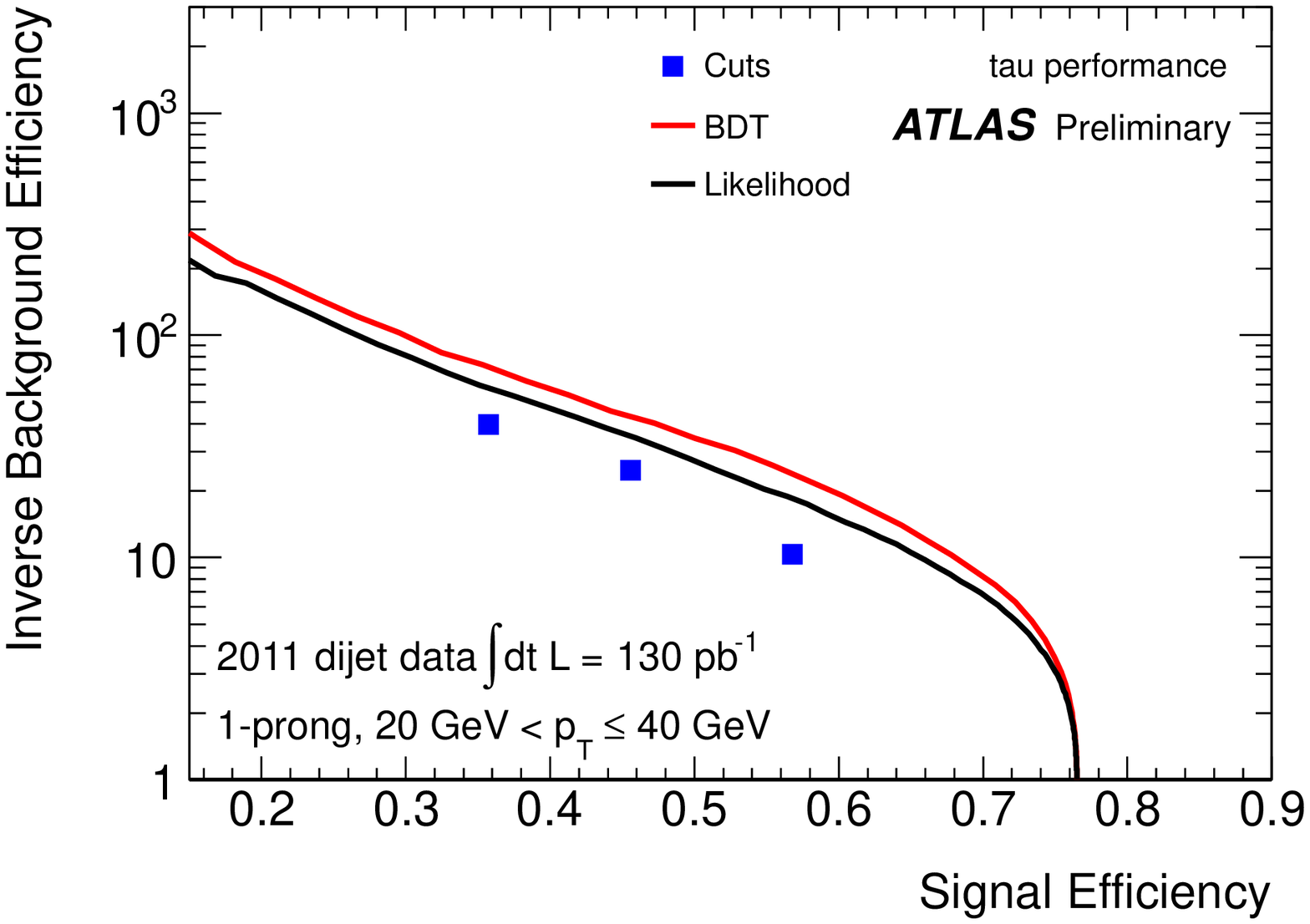} }
\caption{Signal efficiency versus inverse background efficiency for the different tau identification methods shown for 1-prong tau candidates with $20\GeV<\pT<40\GeV$~\cite{ATLAS-CONF-2011-152}.}
\label{fig:Perf1pLowPt}       
\end{figure}

\begin{figure}
\centering
\resizebox{0.75\columnwidth}{!}{ %
  \includegraphics{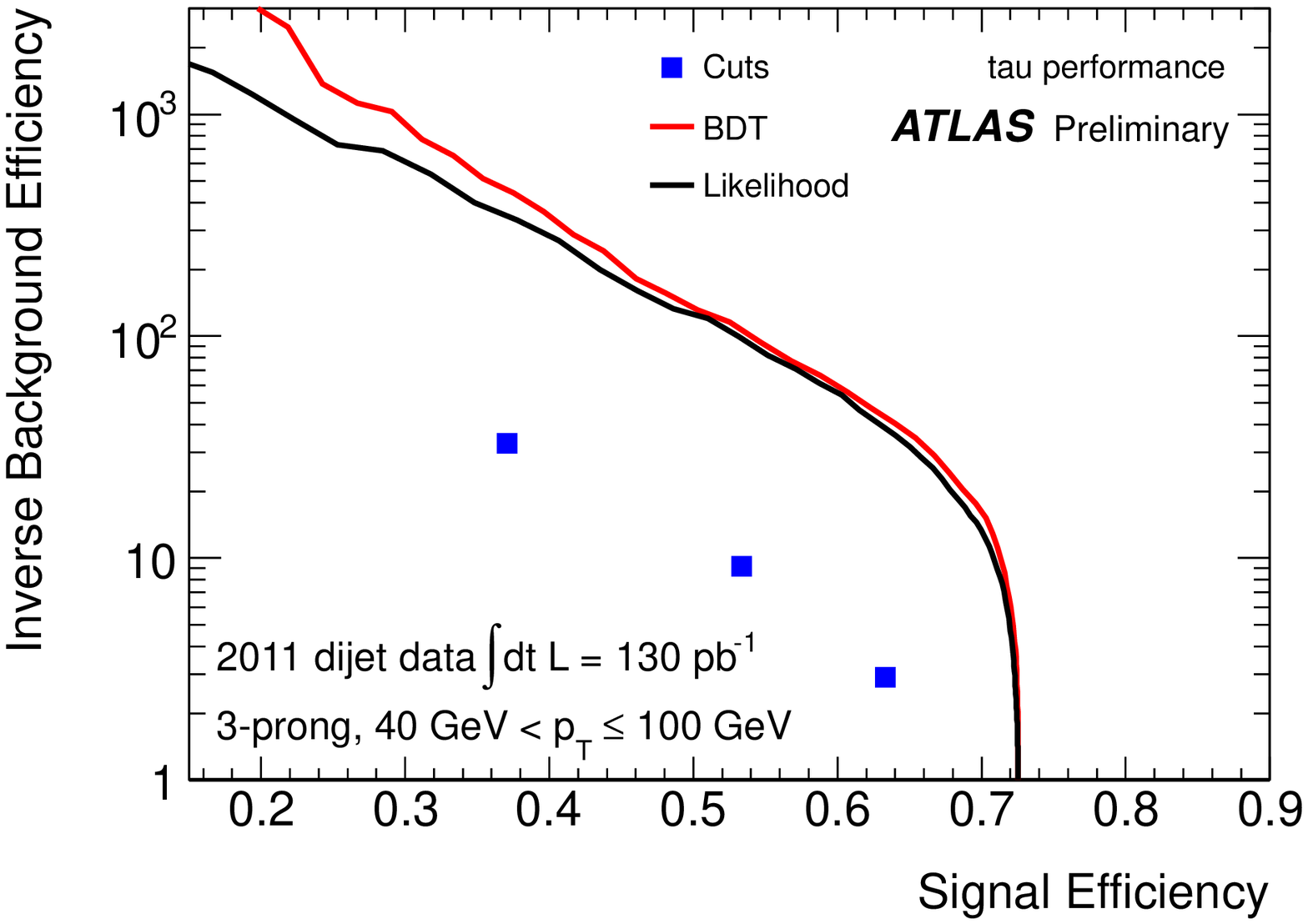} }
\caption{Signal efficiency versus inverse background efficiency for the different tau identification methods shown for 3-prong tau candidates with $40\GeV<\pT<100\GeV$~\cite{ATLAS-CONF-2011-152}.}
\label{fig:Perf3pHighPt}       
\end{figure}

Since there is no attempt to separate QCD jets and tau leptons in the reconstruction process a dedicated identification step is needed. It is based on variables which provide discrimination power between QCD jets and tau leptons. While the charged tracks from the $\uptau$ lepton decay are collimated in a narrow cone, tracks from QCD jets are distributed more widely (Figure~\ref{fig:dRmax3p}). The energy deposit in the calorimeter is also collimated in a small area around the tau axis, while for QCD jets, a larger area is affected (Figure~\ref{fig:calRadius}). 
There are three independent methods for tau identification in ATLAS: a cut-based approach, placing rectangular cuts on variables, a projective likelihood (LLH) method, using the log-likelihood-ratio of signal and background, and boosted decision trees (BDT), to find the optimal separation in a multi-dimensional phase space.
The methods use different sets of identification variables and are separately trained for single- and multi-prong tau candidates. In addition, the likelihood and BDT are trained for different numbers of reconstructed vertices in order to take event pile-up into account. Three dedicated working points with signal efficiencies of $\sim60\%$, $\sim45\%$ and $\sim30\%$ (loose, medium, tight) are provided for all tau identification methods. The likelihood output score is shown in Figure~\ref{fig:Perf3pLLHscore}) for 3-prong tau candidates. For the training of the identification algorithms, the QCD background was obtained from data, while the tau decay signal was simulated in \Wtau\ and \Ztau\ Monte Carlo samples. The inverse background efficiency~\cite{ATLAS-CONF-2011-152} versus signal efficiency for all three methods is shown for 1-prong (Figure~\ref{fig:Perf1pLowPt}) low-\pT\ and 3-prong (Figure~\ref{fig:Perf3pHighPt}) high-\pT\ tau candidates.

\begin{figure}
\centering
\resizebox{0.75\columnwidth}{!}{ %
  \includegraphics{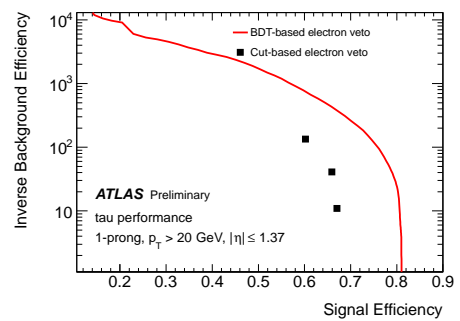} }
\caption{Signal efficiency versus inverse background efficiency for the different tau electron veto methods shown for 1-prong tau candidates with $\pT>20\GeV$ in the central (barrel) part of the detector~\cite{ATLAS-CONF-2011-152}.}
\label{fig:Perf1pEleVeto}       
\end{figure}

Electrons can also be misidentified as a tau lepton. Due to the signature of the electron in the detector, they will be reconstructed mostly as a 1-prong tau-candidate. To distinguish between electrons and such tau leptons two vetoes -- a cut-based and boosted decision tree (BDT)-based -- are available. The performance of these electron vetoes is shown in Figure~\ref{fig:Perf1pEleVeto}. 
\section{Identification Efficiency Measurements }
\label{sec:3}
\begin{figure}
\centering
\resizebox{0.75\columnwidth}{!}{ %
  \includegraphics{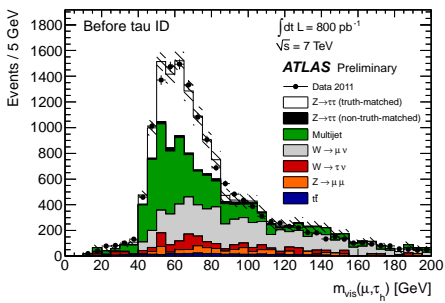} }
\caption{Visible mass of the selected muon and tau candidate for data and Monte Carlo simulation after full event selection, but before applying any tau identification. The QCD multi-jet background was obtained by a data-driven method~\cite{ATLAS-CONF-2011-152}.}
\label{fig:EffZtautauVissmass_before}       
\end{figure}

\begin{figure}
\centering
\resizebox{0.75\columnwidth}{!}{ %
  \includegraphics{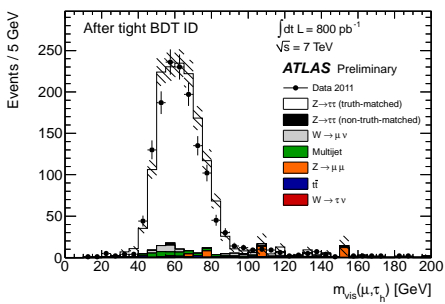} }
\caption{Visible mass of the selected muon and tau candidate for data and Monte Carlo simulation after full event selection and  applying the tight BDT tau identification. The QCD multi-jet background was obtained by a data-driven method~\cite{ATLAS-CONF-2011-152}.}
\label{fig:EffZtautauVissmass_after}       
\end{figure}

The performance and systematic uncertainties of the tau identification methods are evaluated on data using two different signal channels. The first method uses \Ztau\ events in $800\,\ipb$ of ATLAS data and relies on a tag-and-probe approach using the event selection from the ATLAS \Ztau\ cross-section measurement~\cite{Ztautau}. Events are tagged with a muon from a tau decay, and the other tau lepton in the event is required to decay hadronically, forming the probe that is used to measure the identification efficiency. The electro-weak background is dominated by \Wmn\ and was estimated from Monte Carlo simulation, while the QCD multi-jet background was obtained by a data-driven method. The visible mass of the muon and the hadronic tau is shown for data before (Figure~\ref{fig:EffZtautauVissmass_before}) and after (Figure~\ref{fig:EffZtautauVissmass_after}) applying the tight BDT tau identification and agrees well with Monte Carlo predictions.
The tau identification efficiency was also measured using \Wtau\ events collected in $1.37\,\ifb$ of ATLAS data. Variables based on the missing transverse energy were used to select the events. The number of hadronic tau candidates are derived by a template fit of the track multiplicity of the tau candidates. Three different templates were used: real hadronic tau decays, electrons misidentified as tau leptons, and QCD multi-jets misidentified as tau leptons. While the first two are obtained from Monte Carlo simulation, the QCD multi-jet template was estimated from a control region rich in QCD events. The track multiplicity distribution is shown for data and Monte Carlo before (Figure~\ref{fig:EffWtaunu_ntrk_before}) and after (Figure~\ref{fig:EffWtaunu_ntrk_after}) applying the tight BDT tau identification.

The measured efficiencies in both methods are in good agreement with Monte Carlo predictions within 5\% (8\% - 12\%) for the \Wtau\  ($\Ztau\rightarrow\mu\tau_{\mathrm{had}}$ ) method.
\begin{figure}
\centering
\resizebox{0.75\columnwidth}{!}{ %
  \includegraphics{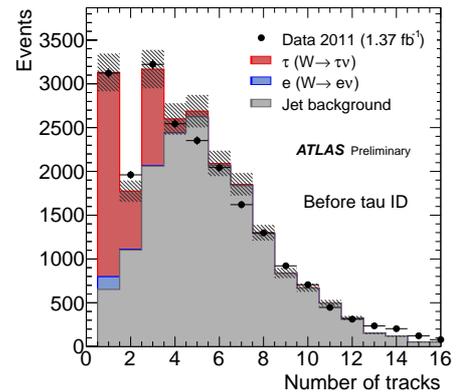} }
\caption{Number of charged tracks for tau candidates after full event selection, but before applying any tau identification. The three different templates are shown~\cite{ATLAS-CONF-2011-152}.}
\label{fig:EffWtaunu_ntrk_before}       
\end{figure}
\begin{figure}
\centering
\resizebox{0.75\columnwidth}{!}{ %
  \includegraphics{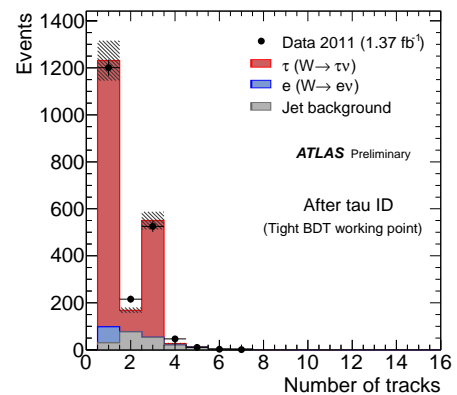} }
\caption{Number of charged tracks for tau candidates after full event selection and applying tight BDT tau identification. The three different templates are shown~\cite{ATLAS-CONF-2011-152}.}
\label{fig:EffWtaunu_ntrk_after}       
\end{figure}
\section{Summary and Conclusion}
\label{sec:Summ}
ATLAS has a large physics program with tau lepton final states, and a well performing tau identification is a essential part of these analyses. Different techniques are used to separate tau leptons from the quark and gluon initiated jet background. The multivariate methods perform better than a simple cut-based approach, especially for tau leptons with a transverse momentum larger than 40~\GeV. The corresponding efficiencies and systematic uncertainties of the tau identification methods have been studied using Standard Model processes. 
%
%
%
%

%
\end{document}
%